\documentclass{aastex}          
\usepackage{spr-astr-addons}    
\usepackage{epsfig}
\usepackage{graphicx}
\usepackage{color}

\begin{document}
%
\title{On the role of transition region on the Alfv\'{e}n wave phase mixing in solar spicules}

\author{Z.~Fazel, and H.~Ebadi\altaffilmark{1}}
\affil{Astrophysics Department, Physics Faculty,
University of Tabriz, Tabriz, Iran\\
e-mail: \textcolor{blue}{z$_{-}$fazel@tabrizu.ac.ir}}

\altaffiltext{1}{Research Institute for Astronomy and Astrophysics of Maragha,
Maragha 55134-441, Iran.}

\begin{abstract}
Alfv\'{e}nic waves are thought to play an important role in coronal heating and solar wind acceleration.
Here we investigate the dissipation of standing Alfv\'{e}n waves due to phase mixing at the presence of
steady flow and sheared magnetic field in the stratified atmosphere of solar spicules. The transition region between chromosphere
and corona has also been considered. The initial flow is assumed to be directed along spicule axis, and the equilibrium magnetic field is
taken 2-dimensional and divergence-free.
It is determined that in contrast to propagating Alfv\'{e}n waves, standing Alfv\'{e}n waves dissipate in time rather
than in space. Density gradients and sheared magnetic fields can enhance damping due to phase mixing. Damping times deduced from
our numerical calculations are in good agreement with spicule lifetimes.
Since spicules are short living and transient structures, such a fast dissipation mechanism is needed to transport their energy to the corona.

\end{abstract}


\section{Introduction}
\label{sec:intro}
The heating mechanism of the solar corona is one of the most mysterious issue in astrophysics \citep{asch04}. In the corona,
the temperature rises to a few million degrees. To maintain such a high temperature in the corona in spite of cooling by heat conduction and radiative
losses, a continuous supply of thermal energy is necessary.
X-ray observations from space experiments (e.g. skylab) have shown that the corona is not uniform and consists of many bright loops.
Two promising models for coronal heating \citep{Erde07} are: heating by small scale flares
triggered by magnetic reconnection, and heating by the dissipation of Alfv\'{e}n waves that propagate in the magnetic flux tubes
\citep{Alf1947, Hollweg1973, Hollweg1986, McKenzie et al.1995}. However, there are other types of waves such as acoustic, slow-mode and fast-mode waves
which are strongly damped or reflected at the steep density and temperature gradients of chromosphere and transition region.
As an origin of Alfv\'{e}n waves, \citet{Kudoh1999} considered a photospheric random motion propagating along an open magnetic flux tube in the solar atmosphere,
and performed MHD simulations for solar spicule formation and the coronal heating. They have shown that Alfv\'{e}n waves transport sufficient energy flux into the corona.
\citet{De2007} estimated the energy flux carried by transversal oscillations generated by spicules and compared by the result of \citet{Kudoh1999}.
They indicated that the calculated energy flux is enough to heat the quiet corona and to accelerate the high-speed solar wind.
There are many proposed mechanisms for the dissipation of Alfv\'{e}n waves. Interaction of Magnetohydrodynamic (MHD) waves with inhomogeneous plasmas in which has density gradients
in both x- and z-directions, can leads to a number of very interesting physical phenomena, such as mode coupling \citep{Moriyasu et al.2004}, resonant absorption
\citep{Ionson1978, Ruderman et al.1997}, magnetohydrodynamic (MHD) turbulence \citep{Matthaeus et al.1999}, and phase mixing \citep{Heyvaerts1983}.
These phenomena can affect the MHD wave dynamics and cause enhanced dissipation and heating of astrophysical plasmas.
The idea is that when a density gradient of a medium is perpendicular to the magnetic field, the Alfv\'{e}n speed is a
function of the transverse coordinate. On each magnetic field line, Alfv\'{e}n waves propagate with their local Alfv\'{e}n speed.
After a certain time and distance, as height increases, perturbations by Alfv\'{e}n waves of neighboring magnetic field lines become
out of phase and have different wavelengths, causing large gradients in Alfv\'{e}n wavefront in the direction of the inhomogeneity,
that is the phase mixing. Dissipation then appears allowing the energy in the wave to heat the plasma.

Phase mixing of propagating Alfv\'{e}n waves in a stratified atmosphere, particularly, in solar spicules has been studied by \citet{Ebadi2012b}.
They concluded that dissipation of propagating Alfv\'{e}n waves can occur in space rather than in time. Moreover, they found that the calculated
damping times are much longer than spicule life times. \citet{Ebadi2013} have studied the phase mixing of Alfv\'{e}n waves in the case of shear flows and fields.
It is determined that the shear flow and field can enhance the damping of standing Alfv\'{e}n waves. \citet{Smith2007} have showed that the enhanced phase
mixing can occur both with density gradients and sheared magnetic fields. Spicules are one of the most fundamental components of the solar chromosphere.
They are seen in spectral lines at the solar limb at speeds of about $20-25$~km~s$^{-1}$ propagating from photosphere into the magnetized low
atmosphere of the sun \citep{Tem2009}. Their diameter and length varies from spicule to spicule having the values from $400$~km to $1500$~km and
from $5000$~km to $9000$~km, respectively.
The typical lifetime of them is $5-15$ min. The typical electron density at heights where the spicules are observed
is approximately $3.5\times10^{16}-2\times10^{17}$ m$^{-3}$, and their temperatures are estimated as $5000-8000$ K \citep{bec68, ster2000}.
\citet{Kukh2006} and \citet{Tem2007} observed their transverse oscillations with the estimated period of $20-55$ and $75-110$ s by analyzing the height
series of $H\alpha$ spectra in solar limb spicules observed.
Recently, \citet{Ebadi2012a} based on \emph{Hinode}/SOT observations estimated the oscillation period of spicule axis around $180$ s.
They concluded that the energy flux stored in spicule axis oscillations is of order of coronal energy loss in quiet Sun. Despite the large
body of theoretical and observational works devoted to the spicules, their ejection mechanism is not clear yet, and observations with high
spatial resolutions are needed to distinguish their origin. \citet{Okamoto2011} investigated the statistical properties of Alfv\'{e}nic waves in solar spicules
based on SOT/Hinode observations. They observed a mix of upward and downward propagating as well as standing waves. They found the occurrence rate of $20\%$
for standing waves. They concluded that in some spicules there are waves propagating upward and downward to form a standing wave in the middle of the spicule.
\citet{Cargill1997} performed the numerical simulations of the propagation of Alfv\'{e}nic pulses in two dimensional magnetic field geometries.
They concluded that for an Alfv\'{e}nic pulse the time at which different parts of the pulse emerge into the corona depends on the plasma density and magnetic
field properties. Moreover, they discussed that this mechanism can interpret spicule ejection forced through the transition region. \citet{Tem2010} studied
the upward propagation of a velocity pulse launched initially below the transition region. The pulse quickly steepens to a shock. They concluded that such a model may
explain speed, width, and heights of classical spicules. This can be a motivation to study the phase mixing of upward propagating Alfv\'{e}n waves.
\\In this paper we are interested to study the effect of transition region between chromosphere and corona on phase mixing.
Section $2$ gives the basic equations and theoretical model. In section $3$ numerical
results are presented and discussed, and a brief summary is followed in section $4$.

\section{Theoretical modeling}
\label{sec:theory}
We consider effects of the stratification due to gravity in $2$D x-z plane in the presence of steady flow and shear field.
The phase mixing and the dissipation of standing Alfv\'{e}n waves are studied in a region with nonuniform Alfv\'{e}n velocity both along
and across the spicule axis.
Non-ideal MHD equations in the plasma dynamics are as follows:
\begin{equation}
\label{eq:mass} \frac{\partial \mathbf{\rho}}{\partial t} + \nabla
\cdot (\rho \mathbf{v}) = 0,
\end{equation}
\begin{equation}
\label{eq:momentum} \rho\frac{\partial \mathbf{v}}{\partial t}+
\rho(\mathbf{v} \cdot \nabla)\mathbf{v} = -\nabla p + \rho
\mathbf{g}+ \frac{1}{\mu_{0}}(\nabla \times \mathbf{B})\times
\mathbf{B}+ \rho\nu\nabla^2\mathbf{v},
\end{equation}
\begin{equation}
\label{eq:induction} \frac{\partial \mathbf{B}}{\partial t} = \nabla
\times(\mathbf{v} \times \mathbf{B})+ \eta\nabla^2\mathbf{B},
\end{equation}
\begin{equation}
\label{eq:divergence} \nabla \cdot \mathbf{B} = 0,
\end{equation}
\begin{equation}
\label{eq:state} p = \frac{\rho RT}{\mu}.
\end{equation}
where $\nu$ and $\eta$ are constant viscosity and resistivity coefficients, $\mu_{0}$ is the vacuum permeability, $\mu$ is the mean molecular weight.
The typical values for $\eta$ in the solar chromosphere and corona are 8 $\times 10^{8}$$T^{-3/2}$ and $10^{9}$$T^{-3/2}$ m$^2$ s$^{-1}$, respectively.
The value of $\rho\nu$ for a fully ionized $H$ plasma is $2.2\times 10^{-17}$$T^{5/2}$ kg m$^{-1}$ s$^{-1}$ \citep{Prie1982}. We assume that spicules
are highly dynamic with speeds that are significant fractions of the Alfv\'{e}n speed. Perturbations are assumed to be independent of y, i.e.:
\begin{eqnarray}
\label{eq:perv}
  \textbf{v} &=& v_{0} \hat{k} + v_{y}(x,z,t) \hat{j}, \nonumber\\
  \textbf{B} &=& B_{0x}(x,z) \hat{i}+B_{0z}(x,z) \hat{k} + b_{y}(x,z,t) \hat{j}
\end{eqnarray}
and the equilibrium sheared magnetic field is two-dimensional and divergence-free as \citep{Del2005,Tem2010}:

\begin{eqnarray}
\label{eq:shear field}
 B_{0x}(x,z) &=& B_{0}e^{-k_{b}z} \cos[k_{b}(x-a)], \nonumber\\
 B_{0z}(x,z) &=& -B_{0}e^{-k_{b}z} \sin[k_{b}(x-a)]
\end{eqnarray}
where $a$ is the spicule radius. Since the equilibrium magnetic field is force-free, the pressure gradient is balanced by the gravity force, which is assumed
to be $\textbf{g}$=$-g\hat{k}$ via this equation:
\begin{equation}
\label{eq:balance}
 -\nabla p_{0}(x,z) + \rho_{0}(x,z) \textbf{g}=0,
\end{equation}
and the pressure in an equilibrium state is:
\begin{equation}
\label{eq:presse}
 p_{0}(x,z)= p_{0}(x)~\exp\left(-\int^{z}_{z_{r}}\frac{dz'}{\Lambda(z')}\right).
\end{equation}
The density profile is written in the following form:
\begin{equation}
\label{eq:density}
\rho_{0}(x,z)= \frac{\rho_{0}(x)T_{0}}{T_{0}(z)}~\exp\left(-\int^{z}_{z_{r}}\frac{dz'}{\Lambda(z')}\right),
\end{equation}
where $\rho_{0}(x)$ is obtained from the Alfv\'{e}n velocity for a phase mixed and stratified atmosphere due to gravity which is assumed to be \citep{De99, Karami2009}:
\begin{equation}
\label{eq:densityx}
 \rho_{0}(x)= \rho_{0} [2+ \tanh(\alpha(x-a))]^{-2},
\end{equation}
and
\begin{equation}
\label{eq:scale}
 \Lambda(z)= \frac{RT_{0}(z)}{\mu g},
\end{equation}
where $\rho_{0}$ is the plasma density at $z=6000$~km, $\alpha$ controls the size of inhomogeneity across the magnetic field. The temperature profile is taken here as a smoothed step function, i.e.:
\begin{equation}
\label{eq:temp}
 T_{0}(z)= \frac{1}{2}T_{c}\left [1+d_{t}+(1-d_{t})\tanh(\frac{z-z_{t}}{z_{\omega}})\right],
\end{equation}
here, $d_{t}=T_{ch}/T_{c}$ which $T_{ch}$ and $T_{c}$ denote the chromospheric temperature at its lower part and the coronal
temperature that is separated from the chromosphere by the transition region. $z_{w}=200$~km is the width of transition region which is located
at the $z_{t}=2000$~km above the solar surface. We put $T_{ch}=15\times10^{3}$~K and $T_{c}=3\times10^{6}$~K.

\begin{figure}
\centering
\includegraphics[width=7cm]{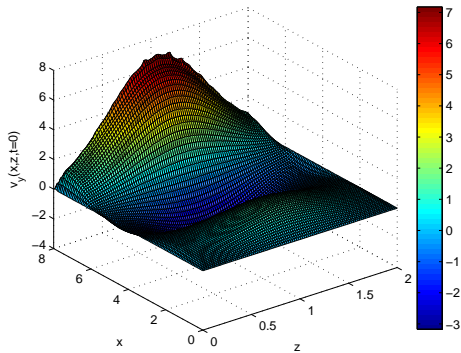}
\caption{(Color online) The initial wave packet for $d=0.5a$ is showed.  \label{fig1}}
\end{figure}

The linearized dimensionless MHD equations with these assumptions are:
\begin{eqnarray}
\label{eq:veloy}
   \frac{\partial v_{y}}{\partial t}  &=& \frac{1}{\rho_{0}(x,z)}\left[B_{0x}(x,z)\frac{\partial b_{y}}{\partial x}+B_{0z}(x,z)\frac{\partial b_{y}}{\partial z}\right], \nonumber\\
   &  & -v_{0}\frac{\partial v_{y}}{\partial z}+ \nu\nabla^{2}v_{y},
\end{eqnarray}
\begin{eqnarray}
\label{eq:mag}
  \frac{\partial b_{y}}{\partial t} &=& \left[B_{0x}(x,z)\frac{\partial v_{y}}{\partial x}+B_{0z}(x,z)\frac{\partial v_{y}}{\partial z}\right] \nonumber\\
   &  & -v_{0}\frac{\partial b_{y}}{\partial z} + \eta\nabla^{2}b_{y},
\end{eqnarray}
where densities, velocities, magnetic field, time and space coordinates are normalized to
$\rho_{\rm 0}$ (the plasma density at dimensionless $z=6$), $V_{A0}$, $B_{\rm 0}$, $\tau$, and $a$ (spicule radius), respectively. Also the gravity
acceleration is normalized to $a^{2}/\tau$. Second terms in the left hand side of Eqs.~\ref{eq:veloy} and ~\ref{eq:mag} present the effect of steady flows.
Eqs.~\ref{eq:veloy}, and ~\ref{eq:mag} should be solved under following initial and boundary conditions:
\begin{eqnarray}
\label{eq:icv}
  v_{y}(x,z,t=0) &=& V_{A0}\exp \left[-\frac{1}{2}(\frac{x-1}{d})^{2}\right]\sin(kz)e^{z/4H} \nonumber\\
  b_{y}(x,z,t=0) &=& 0,
\end{eqnarray}
where $d$ is the width of the initial packet.
\begin{eqnarray}
\label{eq:bc}
  v_{y}(x=0,z,t) = v_{y}(x=2,z,t) = 0, \nonumber\\
  b_{y}(x=0,z,t) = b_{y}(x=2,z,t) = 0, \nonumber\\
  v_{y}(x,z=0,t) = v_{y}(x,z=8,t) = 0, \nonumber\\
  b_{y}(x,z=0,t) = b_{y}(x,z=8,t) = 0.
\end{eqnarray}
Figure~\ref{fig1} shows the initial wave packet given by Eq.~\ref{eq:icv} for $d=0.5a$ ($a$ is the spicule radius).The parameter $k$ is chosen
in such a way to have the standing Alfv\'{e}n wave.

\section{Numerical results and discussion}
\begin{figure}
\centering
\includegraphics[width=8cm]{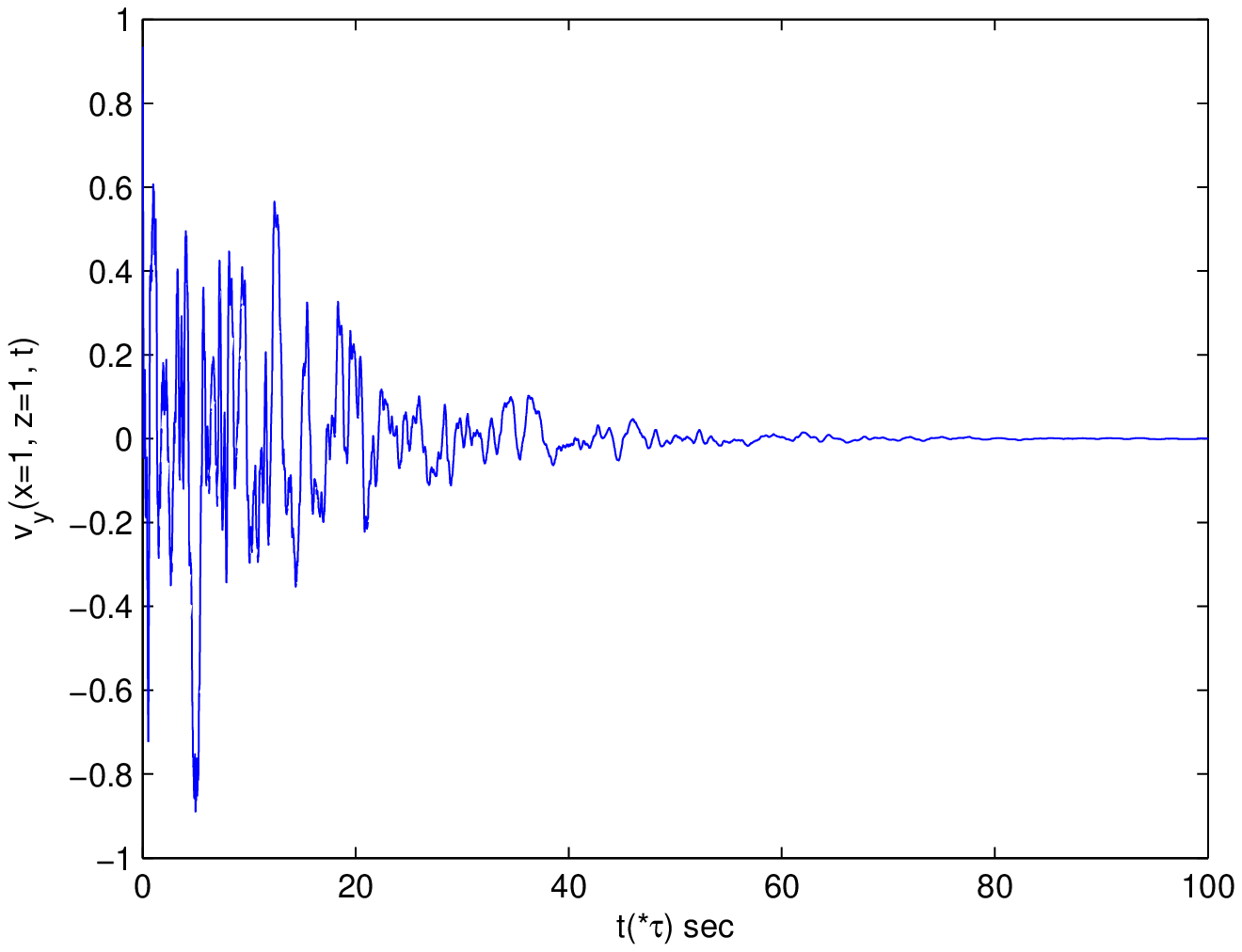}
\includegraphics[width=8cm]{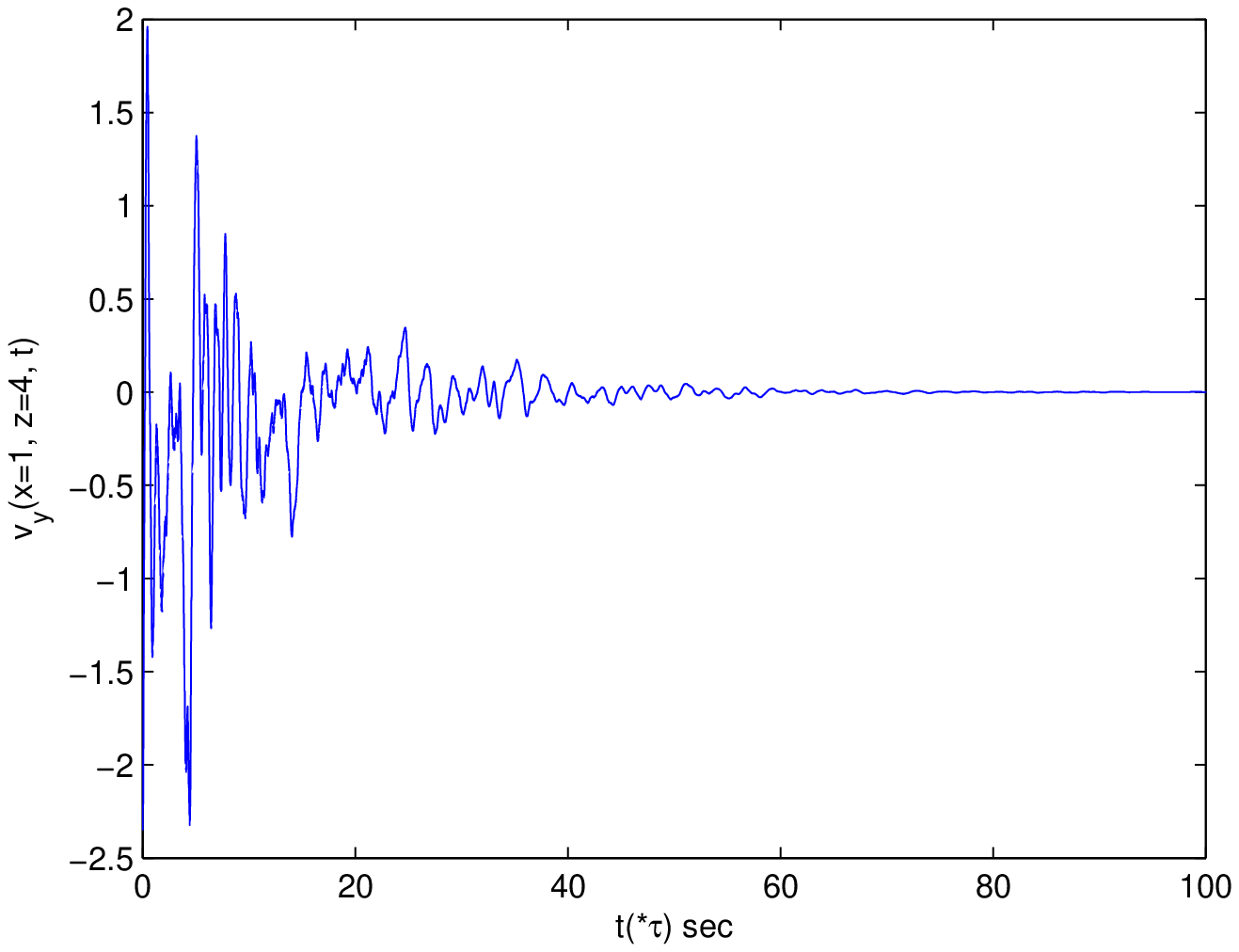}
\includegraphics[width=8cm]{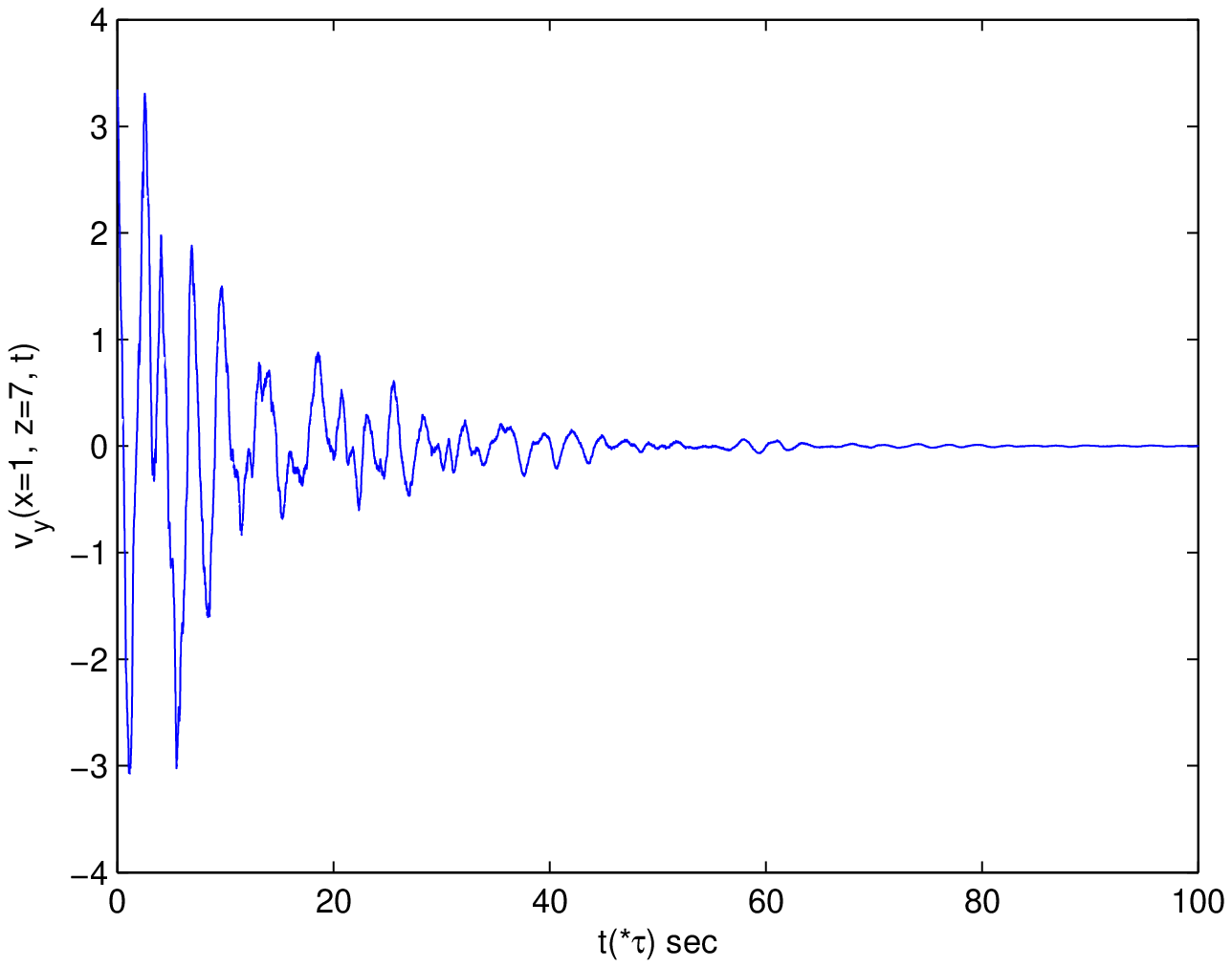}
\caption{(Color online) The perturbed velocity variations are showed with respect to time and $x= 1000$~km for three values of $z= 1000$~km, $z= 4000$~km,
and $z= 7000$~km from top to bottom. \label{fig2}}
\end{figure}

\begin{figure}
\centering
\includegraphics[width=8cm]{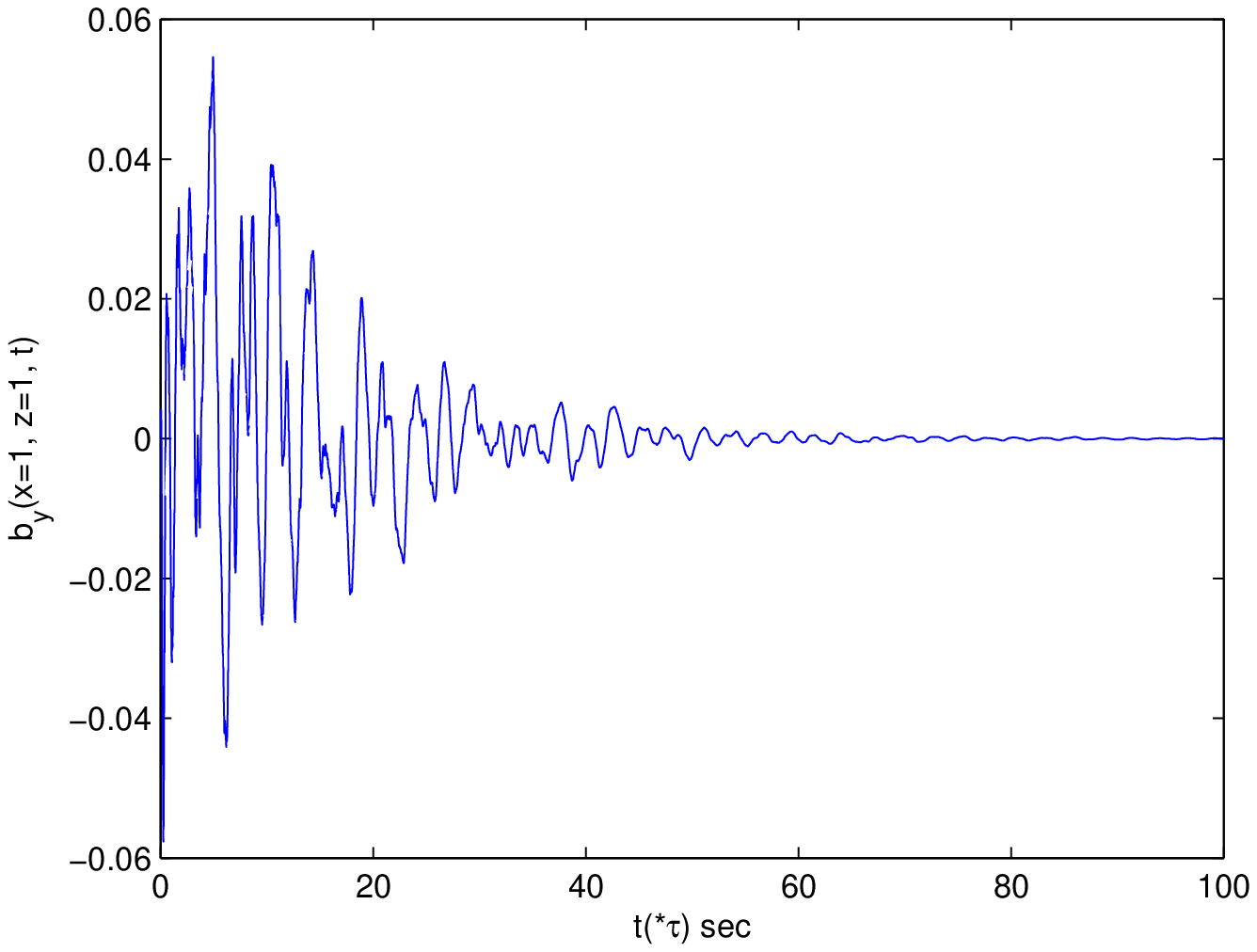}
\includegraphics[width=8cm]{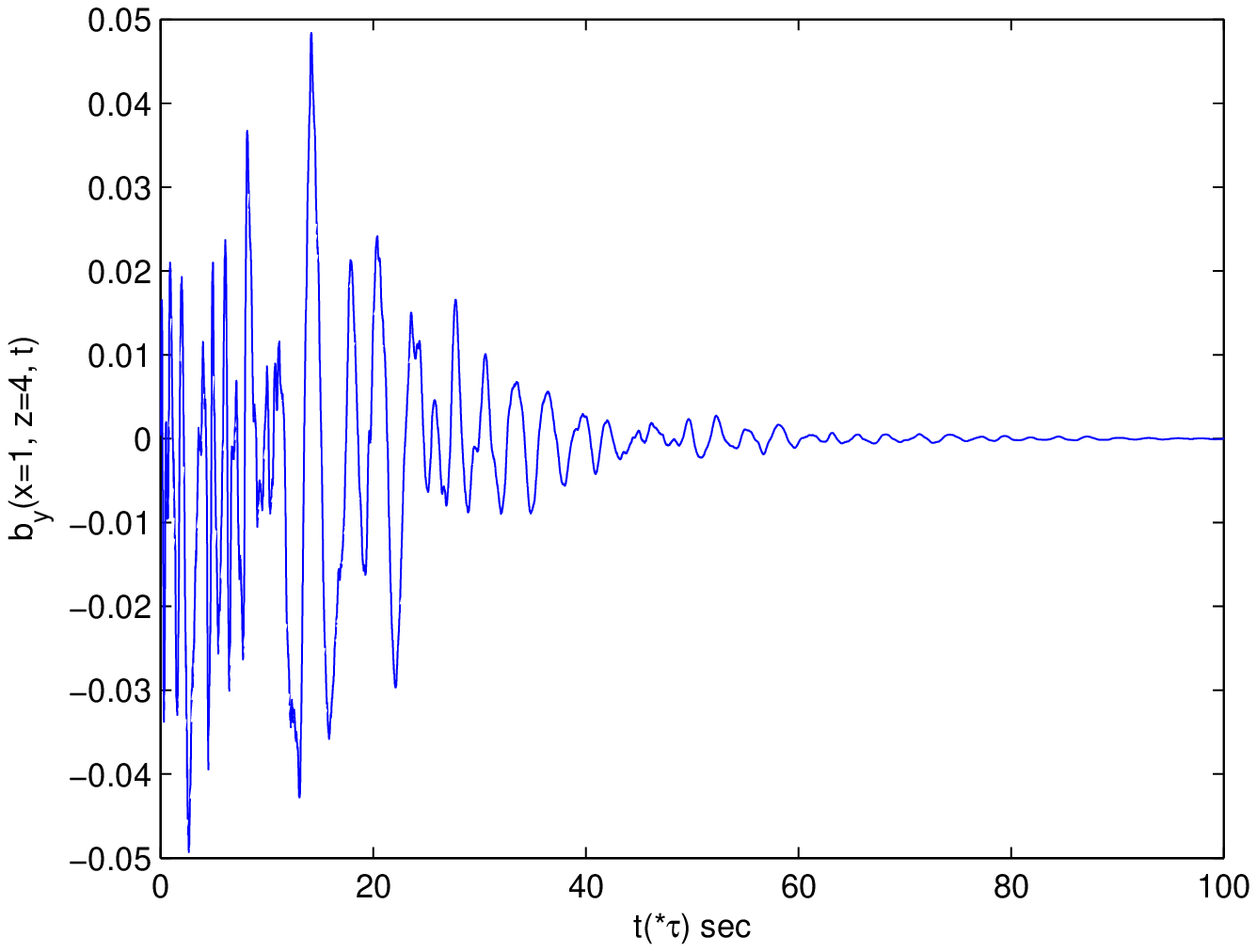}
\includegraphics[width=8cm]{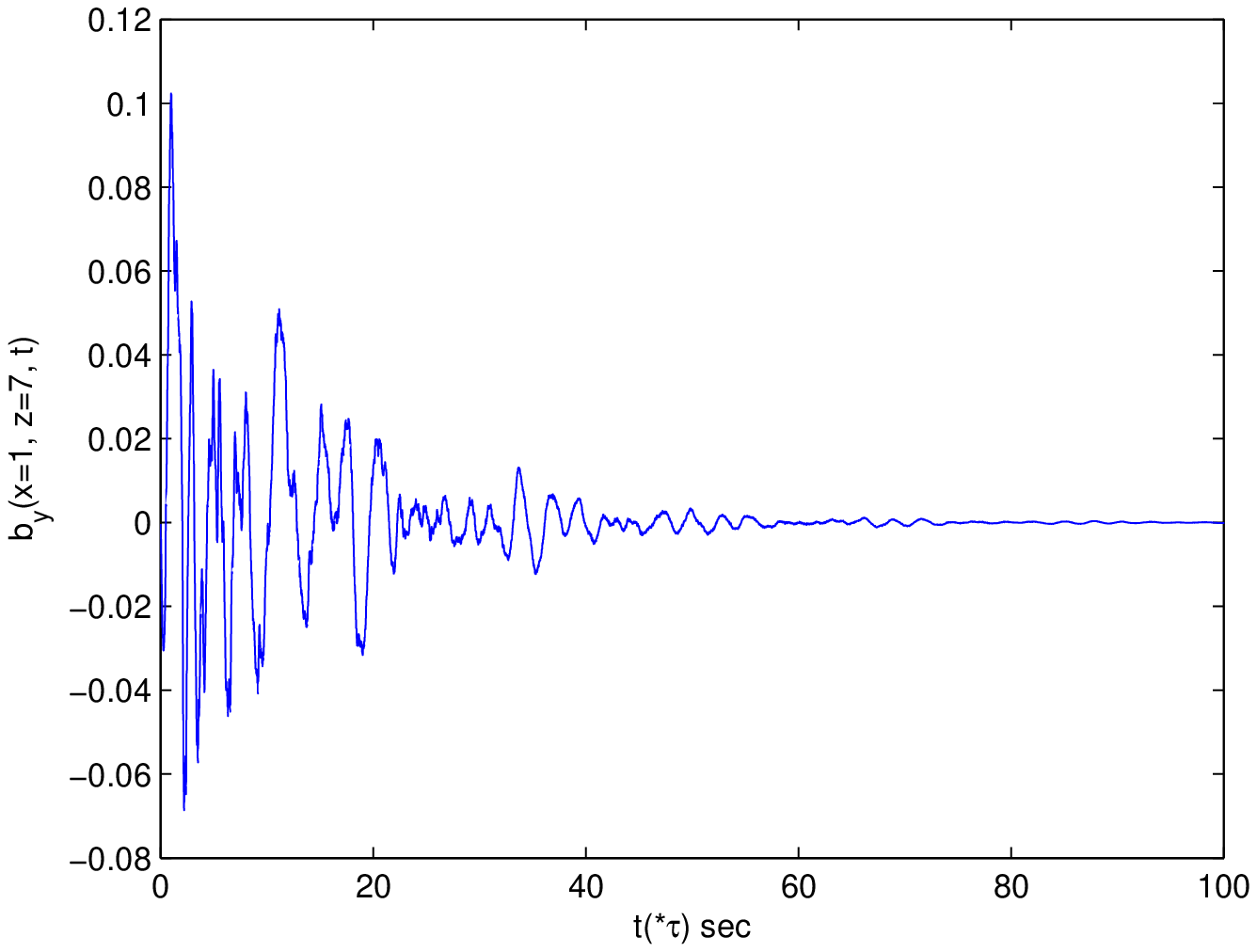}
\caption{(Color online) The perturbed magnetic field variations are showed with the same coordinates as inferred in figure~\ref{fig2}. \label{fig3}}
\end{figure}

To solve the coupled Eqs.~\ref{eq:veloy}, and~\ref{eq:mag} numerically,
the finite difference and the Fourth-Order Runge-Kutta methods are used to take the space and time derivatives, respectively.
We set the number of mesh-grid points as~$256\times256$. In addition, the time step is chosen as $0.0005$, and the system length
in the $x$ and $z$ dimensions (simulation box sizes) are set to be ($0$,$2$) and ($0$,$8$).
The parameters in spicule environment are as follows:
$a=1000$~km(spicule radius), $d=0.5a=500$~km (the width of Gaussian packet), $L=8000$~km (Spicule length), $v_{0}=25$~km s$^{-1}$,
$n_{e}=11.5\times10^{16}$~m$^{-3}$, $B_{0}=1.2\times10^{-3}$~Tesla, $T_{0}=14~000$~K, $g=272$~m s$^{-2}$, $R=8300$~m$^{2}$s$^{-1}$k$^{-1}$
(universal gas constant), $V_{A0}=75$~km/s, $\mu=0.6$, $\tau=13$~s, $\rho_{0}=1.9\times10^{-10}$~kg m$^{-3}$,
$p_{0}=3.7\times10^{-2}$~N m$^{-2}$, $\mu_{0}=4\pi \times10^{-7}$~Tesla m A$^{-1}$, $z_{r}=6000$~km (reference height), $z_{w}=200$~km,
$z_{t}=2000$~km, $x_{0}=1$, $z_{0}=0.5$, $\alpha=2$, $H= 1000$~km, $\eta=10^{3}$~m$^2$ s$^{-1}$, $k_{b}=\pi/8$, and $k= \pi/3$
(dimensionless wavenumber normalized to $a$)\citep{Ebadi2012b}.
\\Figure~\ref{fig2} shows perturbed velocity variations with respect to time in $x=1000$~km, $z=1000$~km; $x=1000$~km, $z=4000$~km;
and $x=1000$~km, $z=7000$~km, respectively. In Figure~\ref{fig3}, perturbed magnetic field variations are presented for $x=1000$~km,
$z=1000$~km; $x=1000$~km, $z=4000$~km; and $x=1000$~km, $z=7000$~km, respectively. In these figures the perturbed velocity and magnetic
field are normalized to $V_{A0}$ and $B_{0}$ respectively, and it is obvious from the plots that there is a damping at the first stage of phase mixing.
This behavior can be related to the presence of transition region between chromosphere and corona.
\\At the first height ($z=1000$~km), total amplitude of both velocity and magnetic field oscillations have values near to the initial ones.
As height increases, the perturbed velocity amplitude does increase in contrast to the behavior of perturbed magnetic field. Nonetheless,
exponentially damping behavior is obvious in both cases. This means that with an increase in height, amplitude of velocity oscillations
is expanded due to significant decrease in density, which acts as inertia against oscillations. Similar results are observed by time-distance
analysis of solar spicule oscillations \citep{Ebadi2012a}. It is worth to note that the density stratification influence on the magnetic field
is negligible, which is in agreement with Solar Optical Telescope observations of solar spicules \citep{Verth2011}.

\begin{figure}
\centering
\includegraphics[width=8cm]{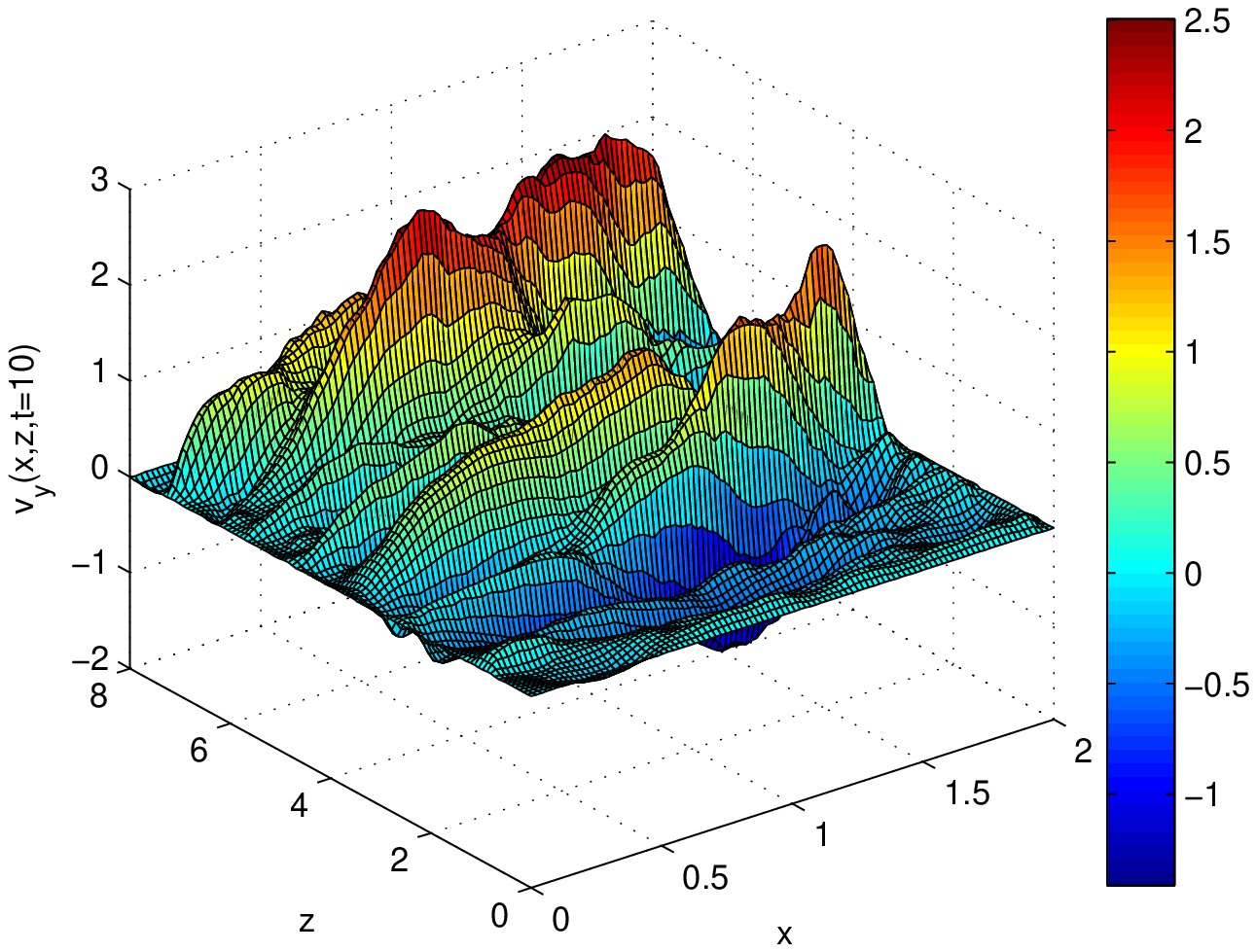}
\includegraphics[width=8cm]{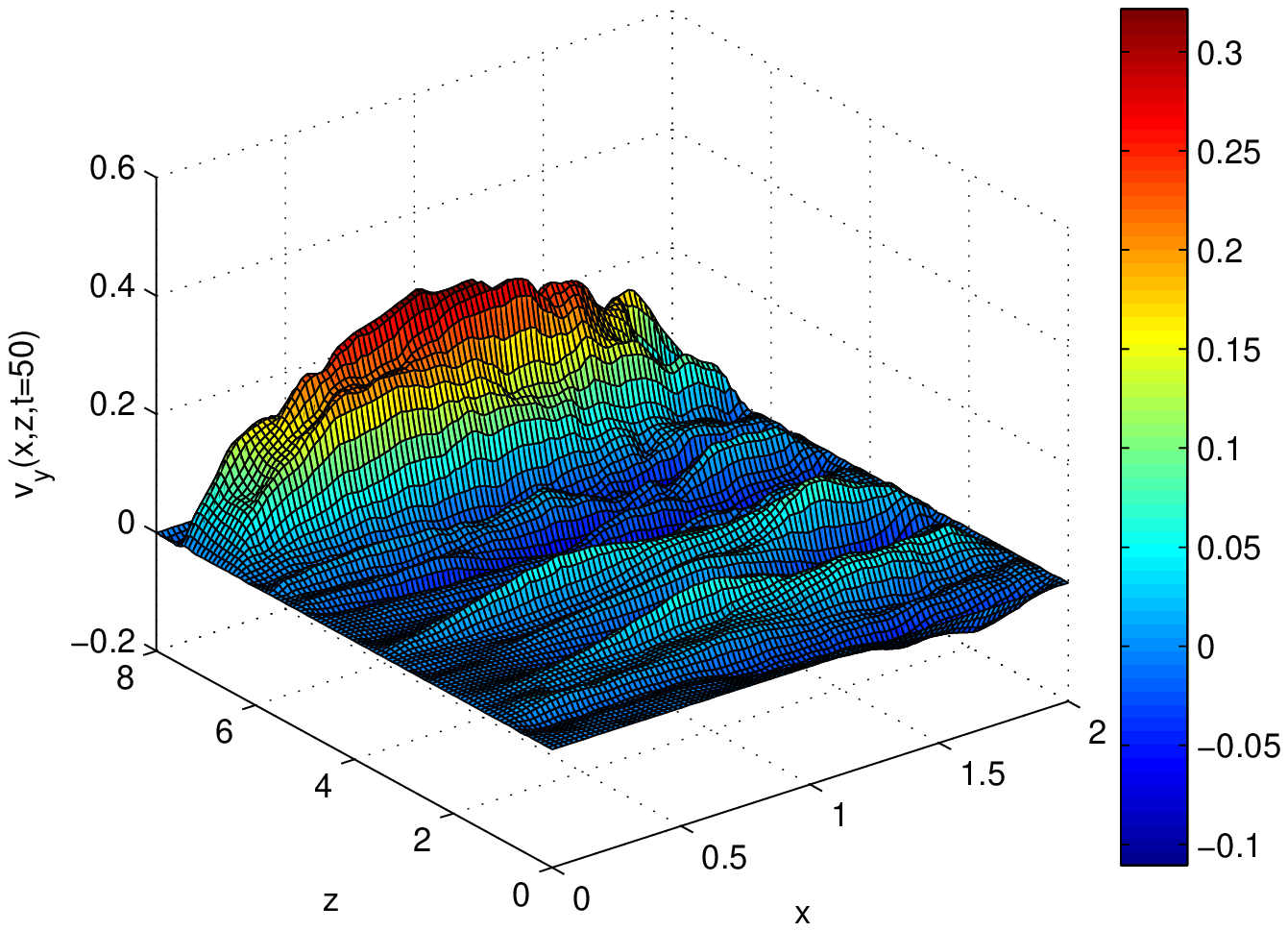}
\includegraphics[width=8cm]{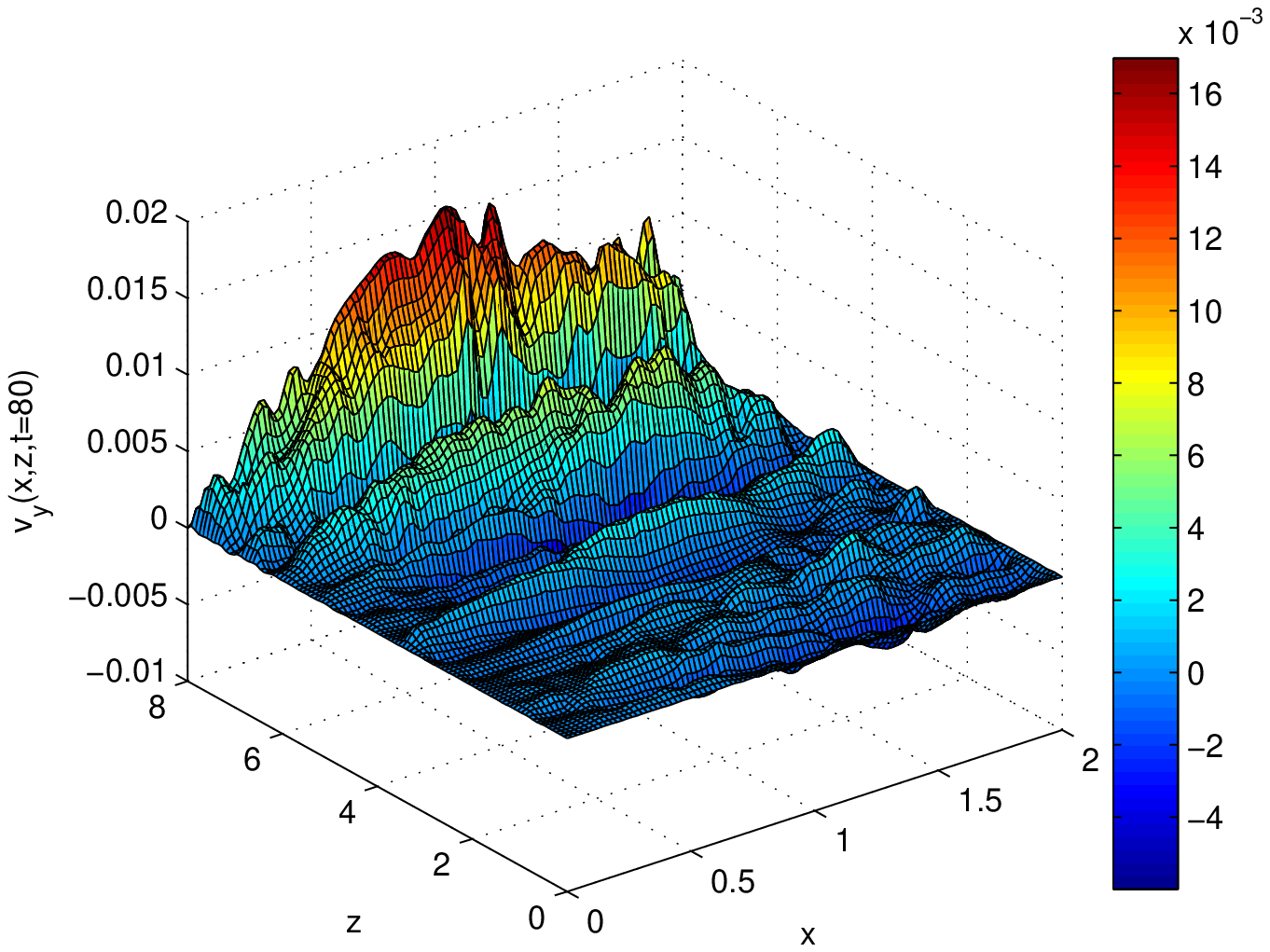}
\caption{(Color online) The $3D$ plots of the transversal component of the perturbed velocity with respect to
$x$, $z$ in $t=10 \tau$~s, $t=50 \tau$~s, and $t=80 \tau$~s for $k_{b}=\pi/8$. \label{fig4}}
\end{figure}

\begin{figure}
\centering
\includegraphics[width=8cm]{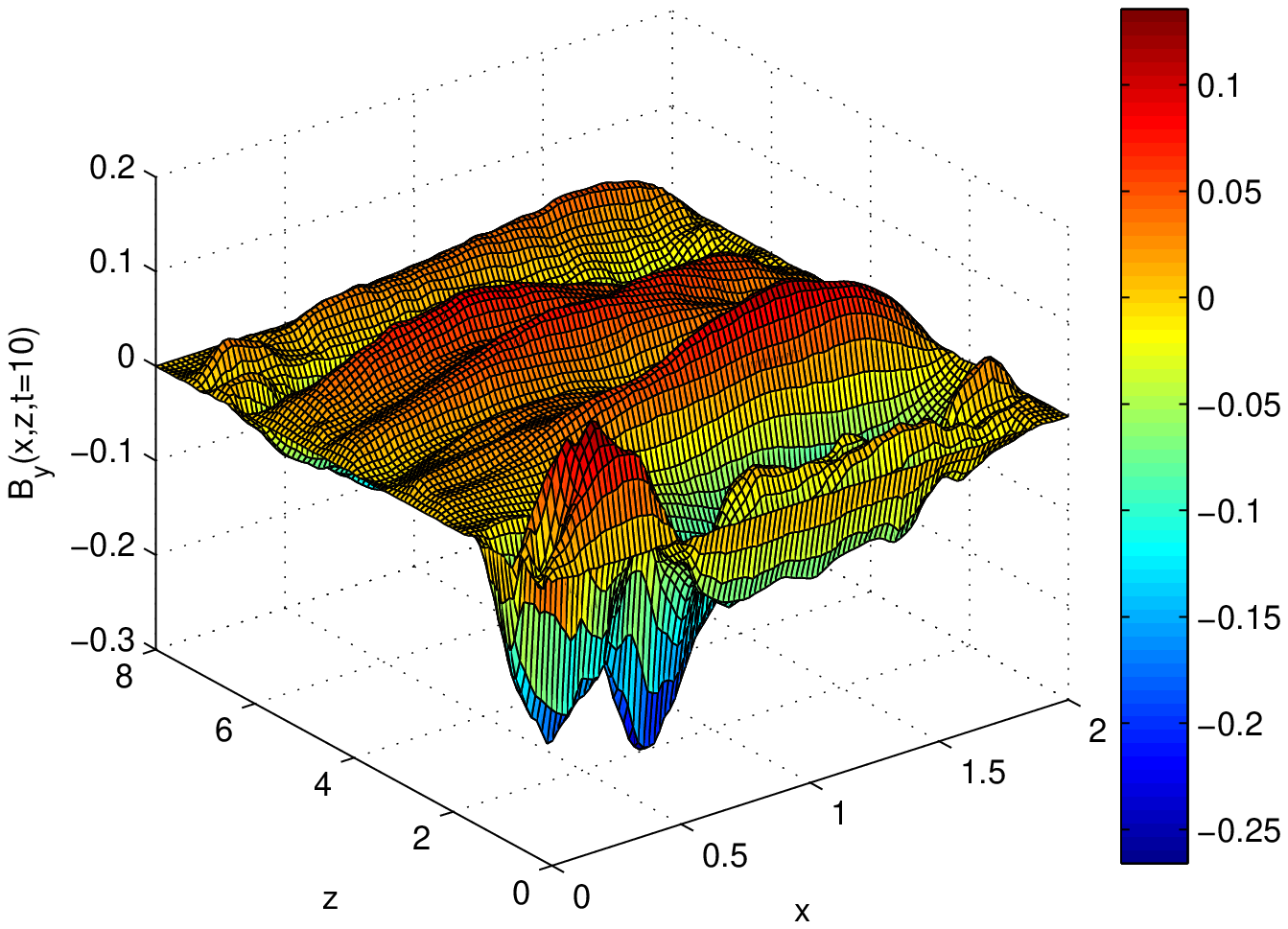}
\includegraphics[width=8cm]{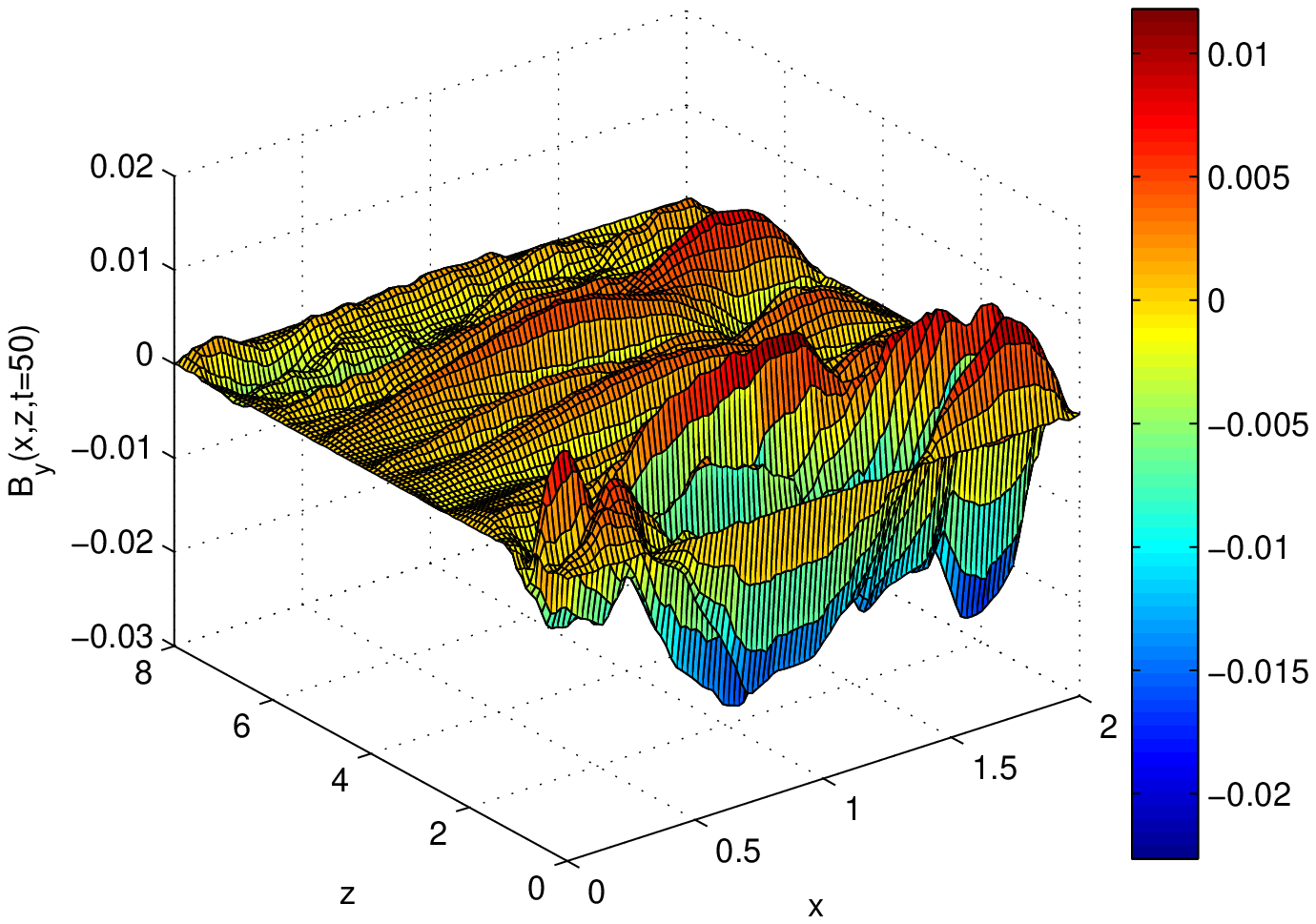}
\includegraphics[width=8cm]{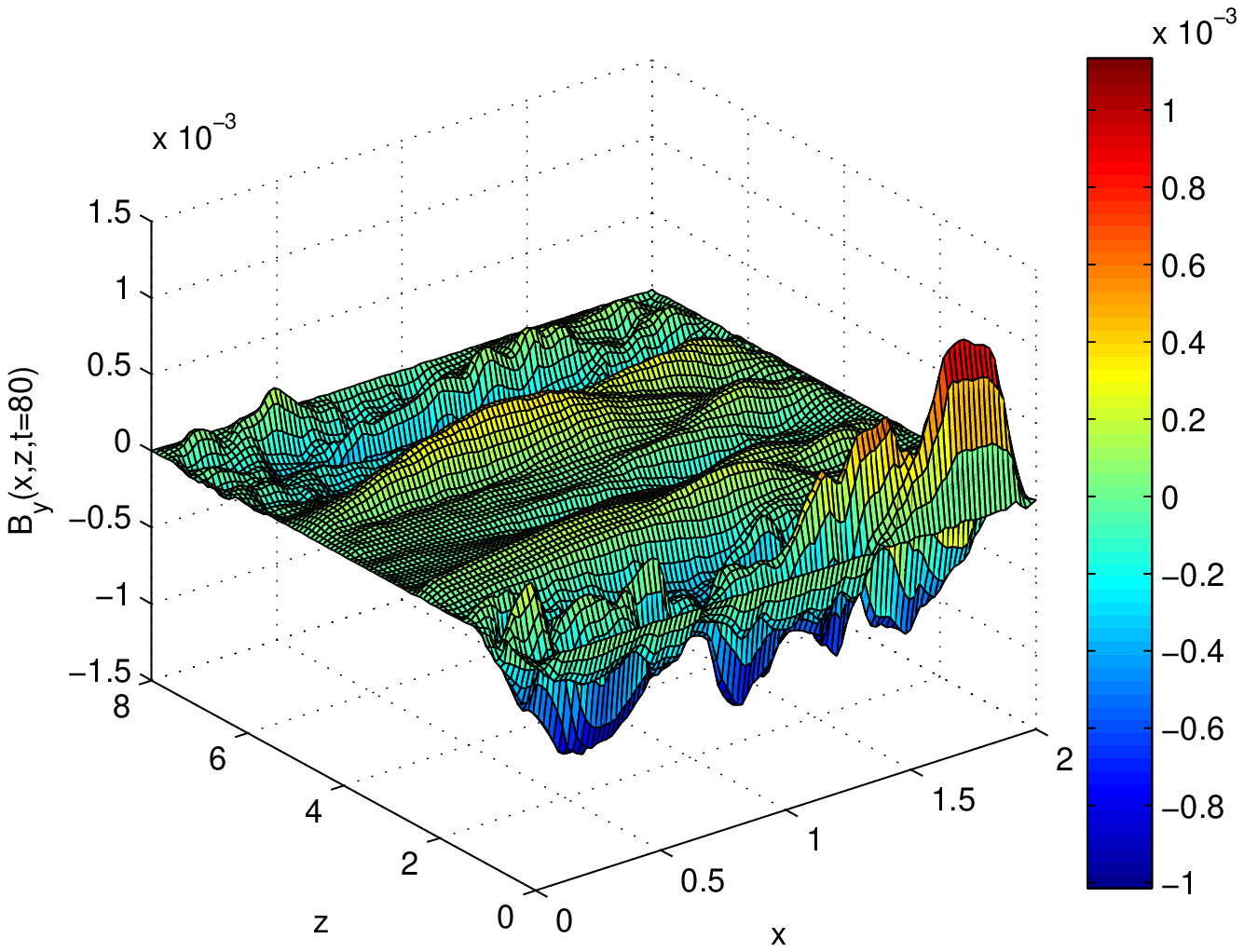}
\caption{(Color online) The same as in Fig.~\ref{fig4} for the perturbed magnetic field. \label{fig5}}
\end{figure}

Figures~\ref{fig4} and \ref{fig5} illustrate the $3D$ plots of the perturbed velocity and magnetic field with respect to $x$, $z$ for $t= 10 \tau$~s,
$t= 50 \tau$~s, and $t= 80 \tau$~s. In the presence of the transition region and stratification due to gravity, the damping process takes place in time than in space.
It should be noted that due to the initial conditions, the damping time scale for the velocity field pattern is longer than the magnetic
field one. In spite of the standing waves considered here, propagating waves are stable and dissipate after some periods due to phase mixing \citep{Ebadi2012b}.

\begin{figure}
\centering
\includegraphics[width=8cm]{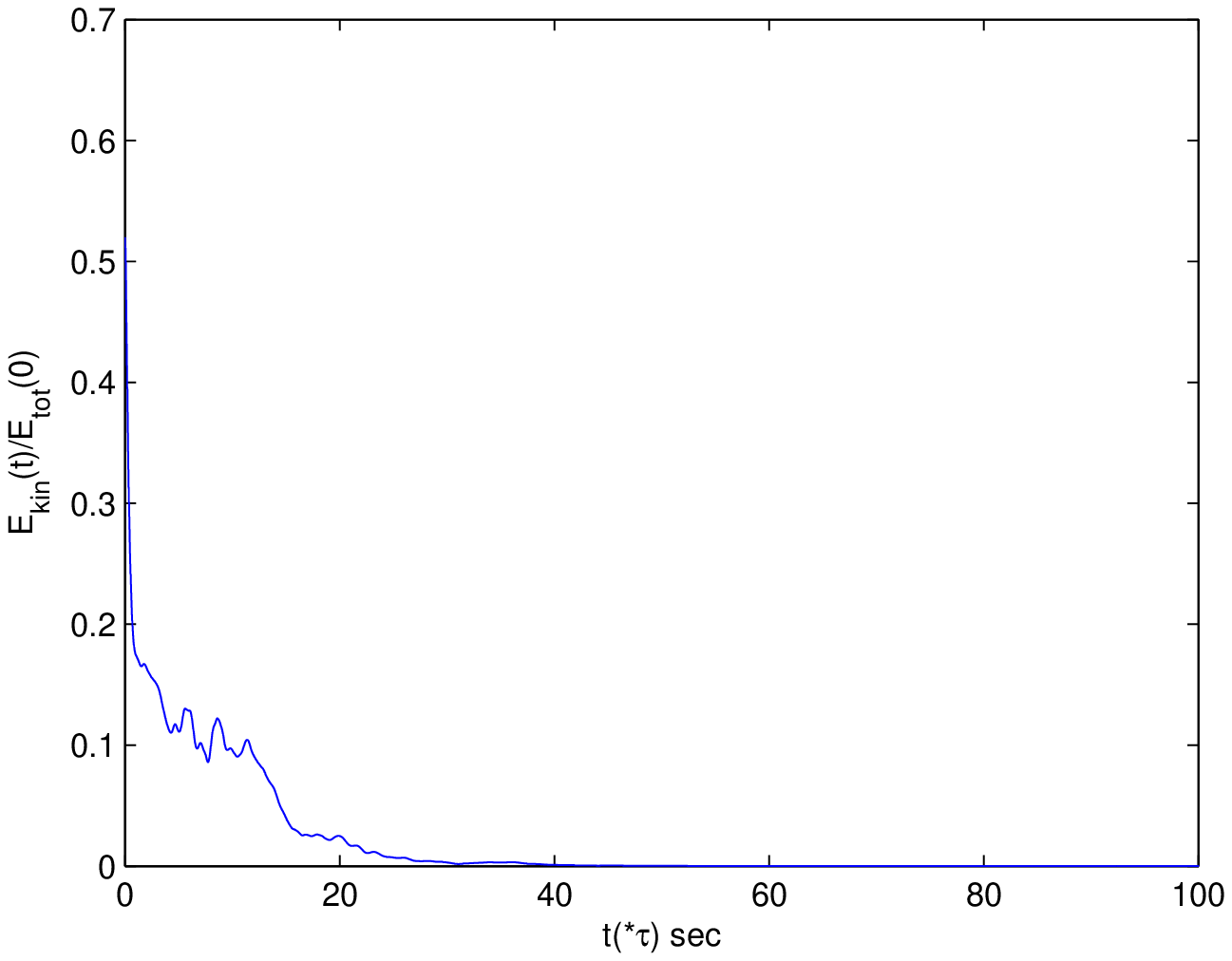}
\includegraphics[width=8cm]{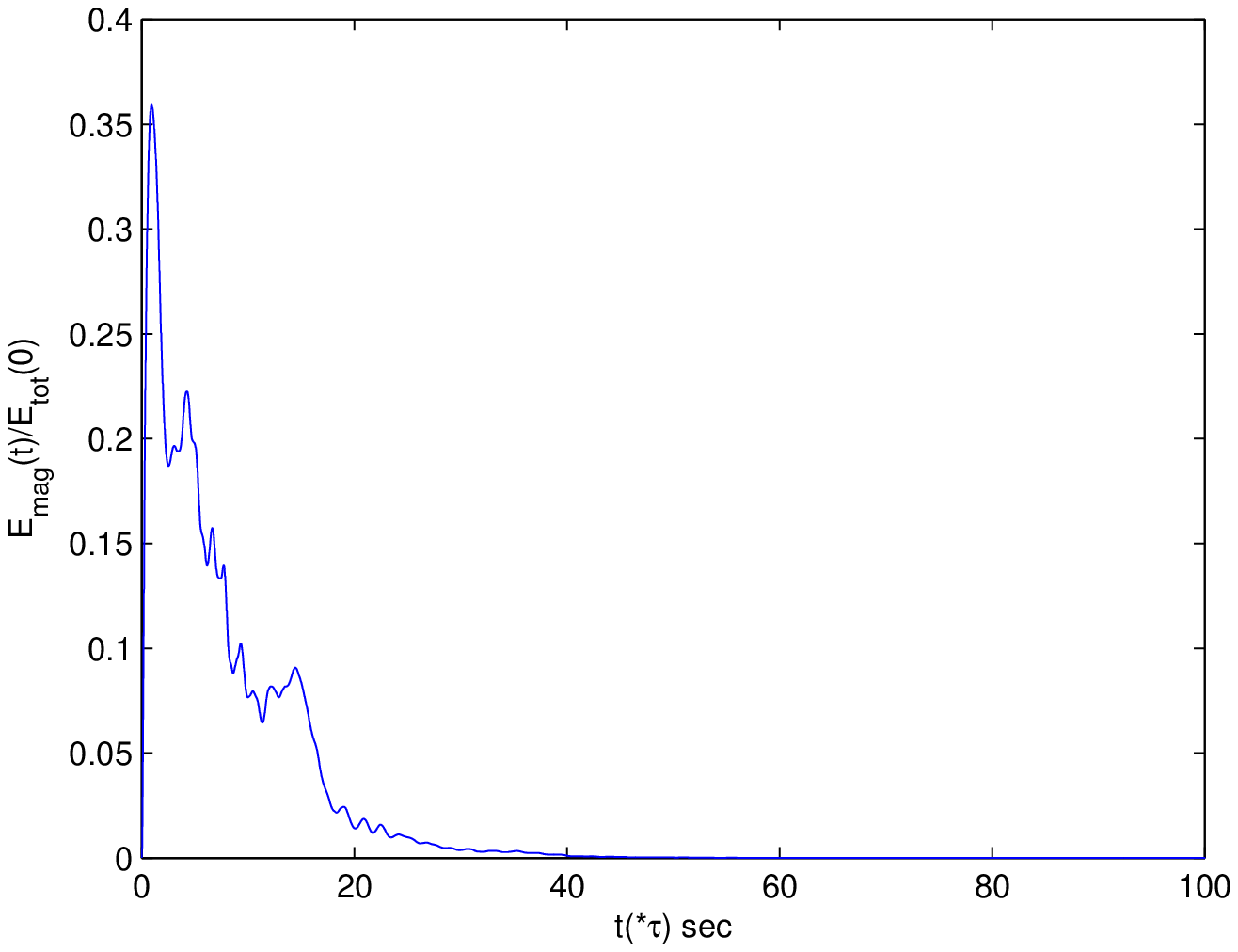}
\includegraphics[width=8cm]{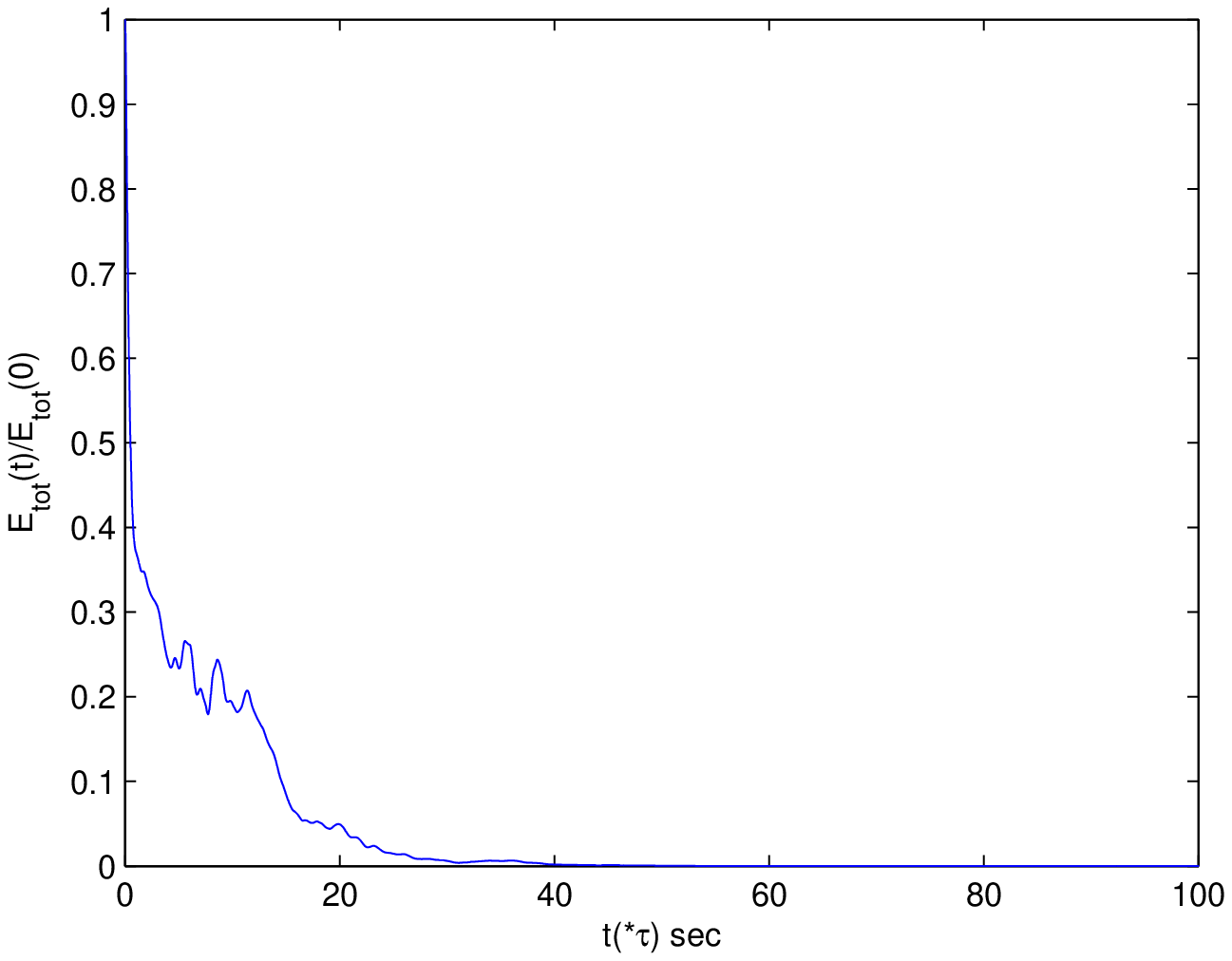}
\caption{(Color online) Time variations of the normalized kinetic, magnetic and total energies for $d= 0.5a$. \label{fig6}}
\end{figure}

In Figure~\ref{fig6}, kinetic, magnetic and total energies normalized to the initial total energy, are presented from top to bottom respectively.
Since spicules have short lifetimes and they are transient phenomena, we claim that in such circumstances the phase mixing can occur in time rather
than in space \citep{De99}. Obtained damping times are in agreement with spicule lifetimes \citep{Tem2009}. Comparison of our results with the results
deduced from the work of \citet{Ebadi2013} shows that density gradients because of transition region make the enhanced phase mixing in standing Alfv\'{e}n waves.
As an example, the total energy amplitude damped to $10\%$ of initial amplitude in $195$ and $325$ seconds, respectively.
This means that the transition region between chromosphere and corona can accelerate damping rates of standing Alfv\'{e}n waves in solar spicules.

\section{Conclusion}
\label{sec:concl}
In spite of propagating Alfv\'{e}n waves, standing Alfv\'{e}n waves can also be found in open structures for example in spicules, and so may
represent an efficient heating mechanism in the solar corona.
In this paper, we assume that spicules are small scale structures in comparison with coronal holes and other mega structures. In our model,
we consider spicules with steady flow and sheared magnetic field. In addition, the transition region between chromosphere
and corona has also been considered. The considered medium is dense in its lower heights while becomes rare and rare as height increases (stratification due to gravity).
Moreover, the density changes across the spicule axis which lead to different Alfv\'{e}n speeds (phase mixing).
Our simulations show that phase mixing can occur in time rather than in space for standing Alfv\'{e}n waves.
The perturbed velocity amplitude increases as height is increased \citep{Ebadi2012a}. In contrast, the perturbed magnetic field amplitude decreases
with height. Both perturbed velocity and magnetic field amplitudes decrease efficiently with time.
Since spicules are short living and transient structures, the fast dissipation mechanism is needed to deliver their energy to the corona.
Our findings illustrate that the transition region can be accounted as a factor that can accelerate damping rates.
We observed that kinetic, magnetic and total energy decrease with time exponentially.

\acknowledgments
This work has been supported financially by the Research Institute for Astronomy and
Astrophysics of Maragha (RIAAM), Maragha, Iran.

\makeatletter
\let\clear@thebibliography@page=\relax
\makeatother

\end{document}